\documentstyle[epsfig]{jaa}
\input{definition.lst}
\begin{document}
\begin{center}
\title[Molecular gas in IRAS-Vela]{Molecular gas associated with the IRAS-Vela shell}
\author[Jayadev Rajagopal \& G. Srinivasan]{Jayadev Rajagopal,\hspace{0.2cm} G. Srinivasan\\Raman Research Institute, Bangalore-560 080, India.\\email: jayadev@stsci.edu, srini@rri.ernet.in}
\date{}
\maketitle
\end{center}
\begin{abstract}

We present a survey of molecular gas in the J = 1$\rightarrow$ 0
transition of $^{12}$CO towards the IRAS Vela shell. The shell,
previously identified from IRAS maps, is a ring-like
structure seen in the region of the Gum Nebula. 
We confirm the presence of molecular gas associated with some of the
infrared point sources seen along the Shell. We have studied the
morphology and kinematics of the gas and conclude that the shell is expanding 
at the rate of $\sim$ 13 \kms from a common center. We go on to 
include in this study the Southern Dark Clouds seen in the region.
The distribution and motion of these objects firmly identify them
as being part of the shell of molecular gas. Estimates of the mass
of gas involved in this expansion reveal that the shell is a massive
object comparable to a GMC. From the expansion and various other signatures
like the presence of bright-rimmed clouds with head-tail morphology, clumpy
distribution of the gas etc., we conjecture that the molecular gas we have
detected is the remnant of a GMC in the process of being disrupted and
swept outwards through the influence of a central OB association, itself 
born of the parent cloud.
\end{abstract}
\section{Introduction}

The IRAS Vela Shell is a ring-like structure seen clearly in the IRAS Sky
Survey Atlas (ISSA) in the 25, 60 and 100 micron maps. This large feature
extending almost 30 degrees in Galactic longitude ($l^{\rm II} = 245^{\circ}$ to
$275^{\circ}$), and discernable from the galactic plane till galactic
latitude $-15^{\circ}$, was first noticed by A.~Blaauw. Subsequently, a
detailed study of this region formed the major part of the thesis by
Sahu (1992). This infrared shell is seen in projection against the Gum
Nebula as a region of enhanced H$\alpha$ emission in the southern part of
the nebula. But it is quite likely that whereas the IRAS shell may be
located in the vicinity of the Gum Nebula, it is unrelated to it. The main
reason for supposing this is that the kinematics of the shell is quite
different from that of the Gum Nebula as a whole. The evidence comes from
emission line studies of this region (e.g., lines of NII). Whereas there
is no conclusive evidence of any expansion of the Gum Nebula, in two 
directions towards approximately
the centre of the shell the emission lines have a ``double-peaked''
structure, consistent with an expansion with a velocity $\sim 10 \pm
2\;\;{\rm km\;s}^{-1}$. Moreover, the Gum Nebula does not appear as
a discernable feature in the IRAS maps.

The shell roughly envelopes the Vela OB2 stellar association (Brandt {\it
et al.} 1971). Two of the brightest known stars, $\zeta$ Puppis (spectral
type 04If) and $\gamma^{2}$ Velorum a Wolf-Rayet binary are also located
close to the shell on the sky. Based on the symmetric location of the
shell with respect to the Vela OB2 association, Sahu argued that this
group of stars is associated with the shell. The distance estimate to Vela
OB2 association is $\sim 450\;{\rm pc}$, and this has been taken as the
distance to the shell as well (Sahu, 1992).

The IRAS Point Source Catalogue (IPSC) also reveals a ring-like structure,
although slightly offset in position from the ISSA shell. From its
emissivity in the infrared, and assuming the standard $1:100$ ratio of
dust to gas, Sahu estimated the total mass of the shell to be $\sim
10^{6}$ solar masses. Presumably much of this mass must be in the form of
molecular gas. The only evidence for this so far is restricted to the
$\sim$ 35 or so cometary globules in the region. These, with head-tail
structures, are distributed in the region of the shell in a manner which
suggests a physical association. From a comprehensive study of these
globules Sridharan (1992a,b) concluded that these small molecular clouds are
expanding about a common centre with a velocity $\sim$ 12 \kms. 
It turns out that this centre of expansion is roughly
centred on the infrared shell as delineated by the IRAS point sources.
This strengthens the case for the cometary globules being associated with
the IRAS Vela Shell. Even so, this would account for only a few thousand
solar masses of molecular gas since the mass of each of the globules is
less than $\sim 100 M_{\odot}$ (Sridharan, 1992b).

\begin{figure}
\vspace{10cm}
\caption[Overall picture of the Gum-Vela region showing the H$\alpha$ emission
as solid lines, the cometary globules as filled circles with tails....]
{Overall picture of the Gum-Vela region showing the H$\alpha$ emission
as solid lines, the cometary globules as filled circles with tails (scaled up
10 times for clarity), and
other important objects in the region. The morphological centre of the system of cometary
globules is indicated by a $+$ sign.  From Sridharan (1992b).}
\end{figure}

The main objective of the investigation reported in this paper was to make
an extensive survey for molecular gas possibly associated with the IRAS shell. A
second objective was to study the kinematics of this gas. For future
reference we show in figure~1 a schematic of the Gum-Vela region, and in
figure~2 the distribution of the IRAS point sources.

\begin{figure}[h]
\begin{center}
\mbox{
\psfig{file=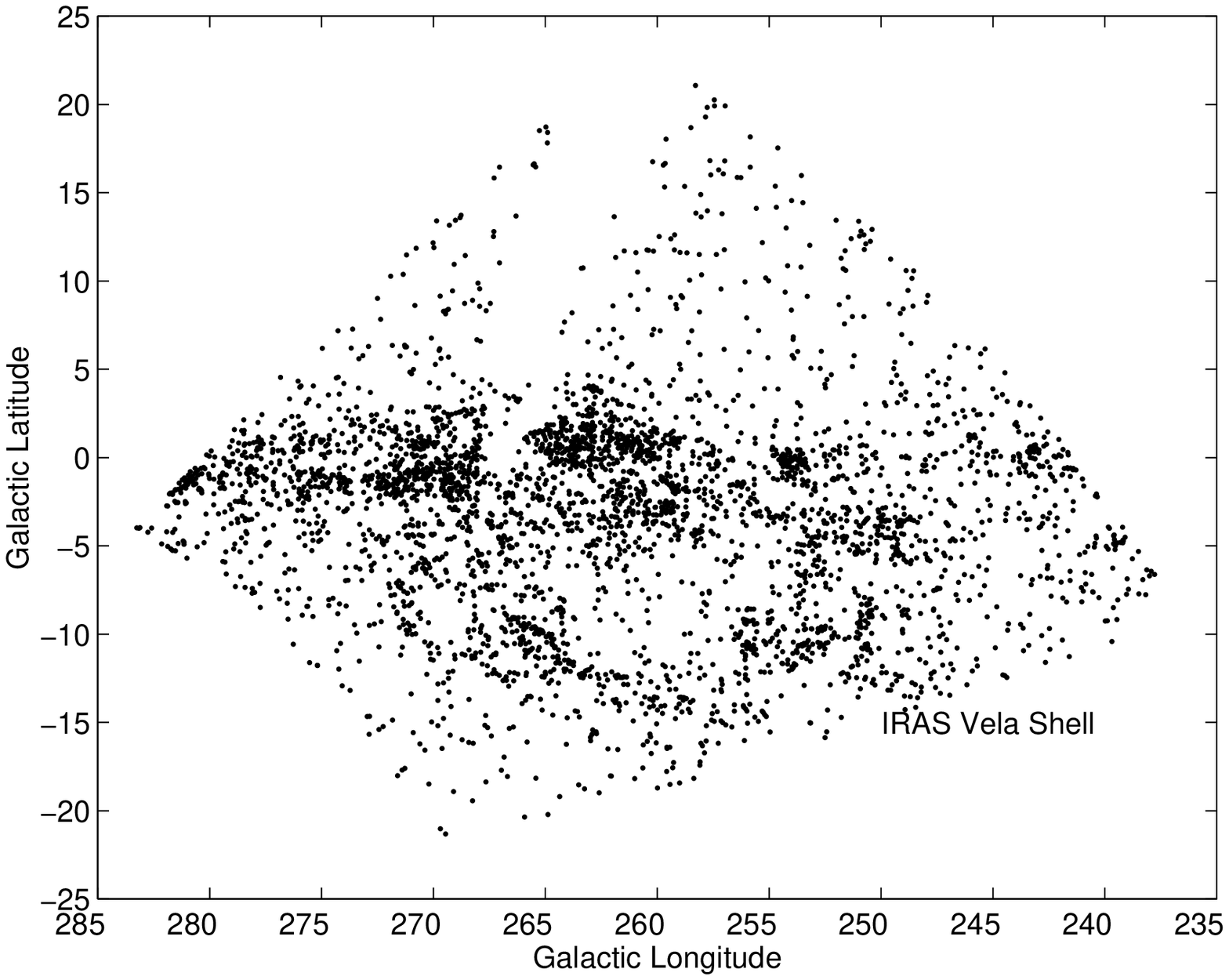,width=9cm,height=9cm}}
\end{center}
\caption[IRAS Point Sources between l = 240$^\circ$ and l = 280$^\circ$.]
{IRAS Point Sources between l = 240$^\circ$ and l = 280$^\circ$. These sources
satisfy the criteria for Young Stellar Object candidates listed in the text. The IRAS
Vela Shell is noticeable below the Galactic plane.}
\end{figure}

\section{Source Selection}

In order to increase our chances of detecting molecular gas we chose a
sample of point sources in the IPSC which were candidates for Young Stellar
Objects (YSO); it had been noted by earlier workers that the shell-like
structure in the distribution of the IRAS point sources was more
pronounced if one restricted oneself to those which are likely to be
associated with YSOs. Prusti (1992), for example, used a certain
``Classifier III'' criteria to pick out the YSO candidates in the IPSC. In
addition to colour and statistical criteria Prusti used certain
``crowding'' properties which tended to enhance the ``shell-structure''.
Since the use of such a filter would excessively bias the distribution of
sources selected for our survey, we used instead a less restrictive
criteria used by Sridharan (1992b), and which are originally due to Emerson
(1987) and Parker (1988). These are listed below:

\begin{enumerate}
\item {Detection at $25 \mu m$ and $60 \mu m$, with $[60 - 25 ] > 0$.}
\item {$[25 - 12] > 0$, if also detected at $12 \mu m$.}
\item {Detection only at $60 \mu m$.}
\item {Detection at $60 \mu m$ and $100 \mu m$ only with $[100 - 60] > 0.6$.}
\item {$[100 - 25] > 0.$}
\end{enumerate}

Here the notation $[25 - 12]$, for example, refers to the {\it 
{\rm 25} to $12 \mu m$ colour ratio}, defined to be $\log \left[S_{25} /
S_{12}\right]$. The flux density in Jansky at $25 \mu m$ is denoted by
$S_{25}$, etc.. We used the filter specified above on the IPSC sources in
the RA range 7 hours to 9 hours which covers the Shell (see figure~2).

\section{The observations} In March--April 1996 we undertook
millimeter-wave observation in the $J = 1 \rightarrow 0$ rotational
transition of the $^{12} {\rm CO}$ molecule at $115.271$ GHz. The
observations were done with the 10.4 m telescope located in the campus of
the Raman Research Institute. It has an altitude-azimuth mount with the
receiver at the Nasmyth focus. The receiver is a Schottky diode mixer
cooled to 20~K. Further details about the telescope and the subsystems may
be found in Patel (1990). The backend used was a hybrid type
correlation spectrometer configured for a bandwidth of 80~MHz with 800
channels giving a resolution of $\approx 100\;{\rm kHz}$ when using both
polarizations. This corresponds to a velocity resolution of $0.26\;\;{\rm
km\;s}^{-1}$.

The observations were done in the frequency switched mode. This had the
advantage that no time is spent looking at source-free regions. Moreover,
since our sources were not point sources ``off-source'' regions are not well
defined for beam switching schemes. A frequency offset of $15.25$ MHz was
chosen between ON and OFF spectra since that is the frequency of the
observed baseline ripple. With this scheme only a polynomial fit was
required to remove any residual baseline curvature. The switching rate was
2~Hz. Calibration was done using an ambient temperature load at intervals
of several minutes. The pointing error was within $20"$ (as determined by
observing Jupiter). The frequency stability of the correlator was checked
by observing the head of the cometary globule CG1 each day. The rms of
this distribution was $0.3\;\;{\rm km\;s}^{-1}$, and hence that will be
the error on the velocities quoted by us. A comparison of the velocities
measured by us towards the heads of several cometary globules with those
measured by Sridharan (1992b) showed good agreement.

\begin{figure}[h]
\begin{center}
\mbox{
\psfig{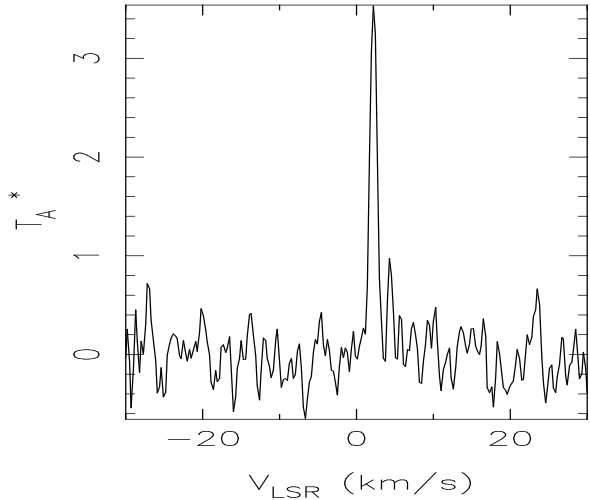}}
\caption[A typical spectrum showing $^{12}$CO emission]{A typical spectrum showing $^{12}$CO emission
(J = 1$\rightarrow$ 0 transition, $\lambda$ = 2.6mm, $\nu_0$ = 115.271 GHz) after removal of a
polynomial fit to the baseline. The spectrum was obtained towards the IRAS point
source IR 56549 in the Shell. The vertical axis shows T$_{\rm A}$$^{\thinspace *}$ (calibrated antenna
temperature) in K. The effective integration time was $\sim$ 20 minutes.}
\end{center}
\end{figure}
Due to the limited observing season we selected about 100 IPSC sources out
of about 3750 which satisfied the various criteria mentioned before. The
sources selected for observations covered the shell although not
uniformly. In addition, we observed in several directions (within the
shell) where there were no IPSC sources. Each observing run consisted of
typically ten minutes of integration. The spectra in the vertical and
horizontal polarizations were then averaged after removing the baseline
curvature. The effective integration time was therefore typically $\sim
20$ minutes. The rms of the noise over such an integration period was
$\sim 0.2\;{\rm K}$. In figure~3 we show a typical spectrum after
correcting for the baseline curvature. The reduction of the data was done
using the UNIPOPS package. The temperature scale  T$_{\rm A}^{\thinspace *}$ is
telescope dependent but we shall not convert it to an absolute scale since
in this paper we are only interested in detection (or otherwise), and the
velocities if molecular material is detected.

\section{Results} We detected $^{12}{\rm CO}$ emission towards 42 of the
100 or so sources observed. Table~1 lists the measured antenna
\begin{table}
\begin{center}
\begin{tabular}{|l|l|l|l|||l|l|l|l|}\hline
Source & V$_{lsr}$ & T$_{\rm A}$$^{\thinspace *}$ & T$_{rms}$ & Source & V$_{lsr}$ & T$_a$$^*$ & T$
_{rms}$\\ \hline\hline
{} & km s$^{-1}$ & K & K &{} & km s$^{-1}$ & K & K\\ \hline
51704&  -1.2&  0.6 & 0.12 &   61322&   8.4&  2.2 & 0.22 \\
52594&  -2.5&  0.9 & 0.18 &   61322&  13.2&  1.5 & 0.22 \\
53668&  -0.5&  0.5 & 0.13 &   61351&   8.1&  1.2 & 0.26 \\
53829&  -3.4&  0.4 & 0.13 &   61428&   7.7&  3.0 & 0.14 \\
54473&  -1.2&  0.8 & 0.18 &   61837&   3.0&  4.6 & 0.20 \\
54599&  -2.2&  3.3 & 0.23 &   62100&   9.9&  7.7 & 0.12 \\
54769&  -3.1&  1.7 & 0.13 &   62578&  14.9&  0.6 & 0.15 \\
55878&  -5.0&  4.5 & 0.13 &   62717&   3.9&  3.3 & 0.21 \\
55884&   4.4&  2.8 & 0.18 &   62841&  -1.0&  2.1 & 0.45 \\
55925&   5.5&  0.9 & 0.24 &   62847&   5.4&  0.7 & 0.17 \\
55932&  21.5&  0.8 & 0.14 &   63338&  14.4&  0.7 & 0.17 \\
56549&   2.2&  3.6 & 0.27 &   63338&   9.4&  1.9 & 0.17 \\
56831&  -1.5&  1.2 & 0.17 &   63927&   4.6&  1.4 & 0.20 \\
57035&  16.4&  1.5 & 0.16 &   64029&   8.8&  3.7 & 0.16 \\
57035&  28.6&  1.9 & 0.16 &   64154&  14.1&  1.4 & 0.16 \\
57898&   1.4&  5.2 & 0.37 &   64388&   0.9&  2.2 & 0.14 \\
58784&  -8.2&  1.1 & 0.18 &   64728&  10.6&  1.2 & 0.27 \\
58793&   8.6&  0.3 & 0.12 &   64999&  -4.0&  1.1 & 0.25 \\
58793&   9.6&  0.3 & 0.12 &   65728&   7.0&  2.4 & 0.27 \\
59579&  -7.5&  0.9 & 0.24 &   65877&   4.9&  9.3 & 0.31 \\
59584&   5.7&  0.8 & 0.18 &   66001&   1.8&  1.3 & 0.36 \\
59891&  -6.3&  0.9 & 0.24 &   67910&   6.0&  3.3 & 0.30 \\
60933&   0.3&  1.3 & 0.18 &   69185&   3.5&  2.0 & 0.30 \\ \hline
\end{tabular}
\caption[Summary of molecular detections toward IRAS Point Sources]
{Summary of molecular detections toward IRAS Point Sources: Column
1 lists the IRAS Point Source Catalogue number for the source. Columns 2
and 3 give the LSR velocity and the corrected antenna temperature (see section 2.5).
Column 4 gives the rms noise level for each source. {\bf The table is continued in
Columns 5 to 8.}}

\end{center}
\end{table}
temperatures and LSR velocities found by fitting gaussians to the spectra.
In some cases multiple features were detected at different velocities. The
distribution of the observed sources in galactic coordinates is shown in
figure~4; the circles denote detection of CO emission and the crosses
indicate non-detections. While the distribution of the open circles
suggests a ring-like structure it is not convincing. A gap in the
distribution of detections in the lower right hand side of figure~4 is
conspicuous. Interestingly, a number of cometary globules are located in
this region. They, however, were not included in our sample selected from
the IPSC. Since molecular gas has earlier been detected in these globules,
their inclusion would fill in this gap (as we shall see in figure 5).
\begin{figure}[h]
\begin{center}
\mbox{
\epsfig{file=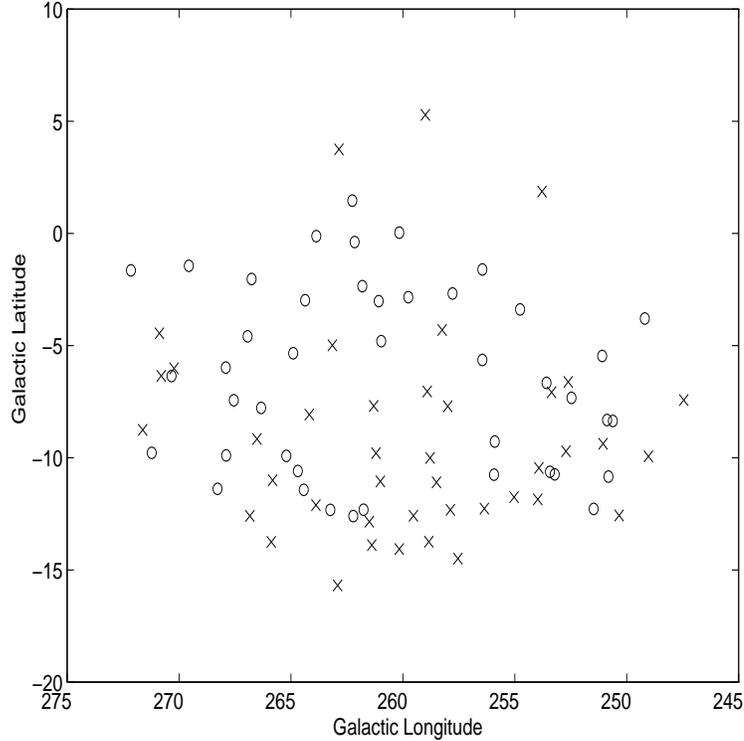,width=10cm,height=10cm}}
\end{center}
\caption[Sources observed by us for CO emission are shown in Galactic coordinates.]
{IPSC sources observed by us for CO emission are shown in Galactic coordinates. The
circles denote detections and the crosses indicate non-detections.
 We have 42 detections from $\sim$ 100 pointings. The sensitivity
limit was $\sim$ 0.6 K.}
\end{figure}

\section{The kinematics of the molecular gas} In this section we wish to
present an analysis of the kinematics of the molecular gas detected by us
in the region of the IRAS Vela Shell. As already mentioned, there are two
independent pieces of evidence to suggest that the shell-like structure
may be in a state of expansion. To recall, the ionized gas possibly
associated with the IRAS shell, as well as the cometary globules, both
show evidence of expansion with a velocity $\sim 12\;\;{\rm km\;s}^{-1}$.
If this is a general feature of the region under discussion then one would
expect the more widely distributed molecular gas also to be in a state of
expansion. Our analysis confirms this.

Sridharan's study revealed that the system of cometary globules are
expanding with the centre of expansion roughly coincident with the
``morphological centre'', i.e., the point towards which the tails of the
maximum fraction of the globules extrapolate to. This is located roughly
at $l = 260^{\circ}$ and $b = -4^{\circ}$. It is reasonable to assume that
if the molecular gas detected by us is also expanding then it is likely to be
with respect to the same ``centre''. The analysis presented below is
predicated on this assumption.

\begin{figure}
\begin{center}
\epsfig{file=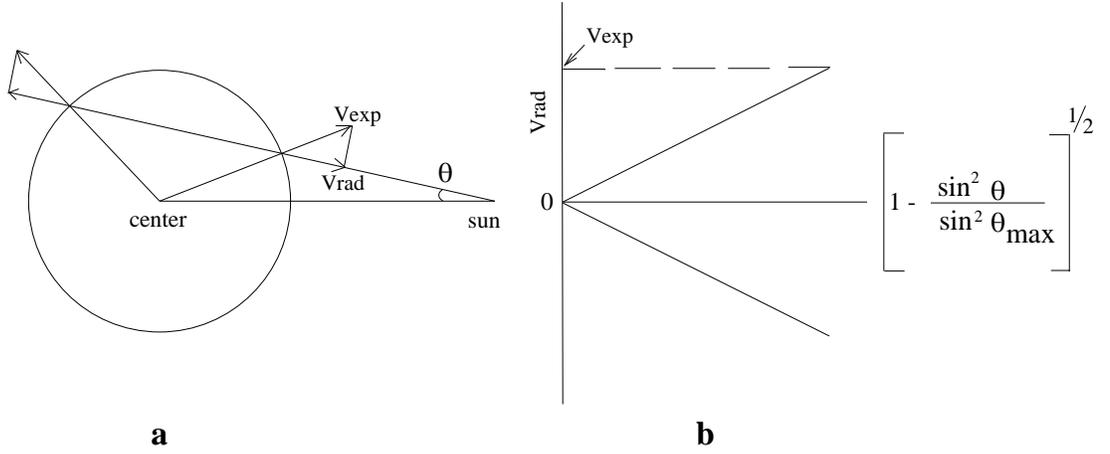,width=14.5cm}
\end{center}
\caption[(a) A schematic for deriving expected radial velocities
from an expanding shell. $V_{exp}$ is the expansion velocity, $V_{rad}$...]
{(a) A schematic for deriving expected radial velocities
from an expanding shell. $V_{exp}$ is the expansion velocity, $V_{rad}$
is the radial component and $\Theta$ is the angular separation of any
point on the shell from the center. (b) $V_{rad}$ plotted against
(1$-sin^2\theta/sin^2\theta_{max})^{1/2}$.}
\end{figure}

Figure~5 illustrates the analysis procedure. The first step is to remove
the contribution to the observed radial velocity due to the differential
rotation of the Galaxy. Since we are looking for a possible expansion with
respect to a common centre, we subtracted from the measured LSR velocity
of each detection the radial velocity component
at the assumed centre of expansion (viz. the morphological centre
of the system of cometary globules) due to galactic differential rotation. 
The radial component of the rotation
velocity was determined from the well known relation \[ v_{r} = {\rm Ad} \sin 2l \cos^{2} b \]

As already mentioned,the co-ordinates of the
assumed centre of expansion are $l = 260^{\circ}$, $b = -4^{\circ}$. Following Kerr and Lynden-Bell (1986) we
assumed a value of $14.5\;\;{\rm km\;s}^{-1}\;\;{\rm kpc}^{-1}$ for Oort's
constant $A$. As for the heliocentric distance $d$ to the centre of the
shell, we adopted a value of 450~pc. This is consistent with the recent
distance estimate to the Vela OB2 association based on the data from
Hipparcos (de Zeeuw {\it et al.}, 1997). It is also consistent with the
distance estimate to the young star embedded in the head of the cometary
globule CG1 (Brand {\it et al.}, 1983).

It may be seen from figure~5 that if the objects expanding about the
common centre with a velocity $v_{\rm exp}$ are distributed on a thin {\it
shell} with a hollow interior then the residual radial velocity $v_{\rm res}$
(i.e., after
allowing for galactic rotation) will be related to the
expansion velocity by \[ v_{\rm res} = \pm \;v_{\rm exp} \left(1 -
\sin^{2} \theta / \sin^{2} \theta_{\rm max} \right)^{\frac{1}{2}} \]

\noindent where $\theta$ is the angular separation of the object from the centre of
expansion, and $\theta_{\rm max}$ is (half) the angular size of the shell
($\sim 12.5^{\circ}$ in the present case). If the sources are distributed on
a thin shell then in a plot of the residual
radial velocity {\it versus} $\left(1 - \sin^{2} \theta / \sin^{2}
\theta_{\rm max}\right)^{\frac{1}{2}}$ the points would lie along two
straight lines as shown in figure~5. If, on the other hand, the objects in
question were distributed over the whole expanding volume then the points
would lie {\it within} the envelope defined by the two lines (provided, of
course, the inner objects are moving slower than the outer ones).

\begin{figure}[h]
\begin{center}
\mbox{
\epsfig{file=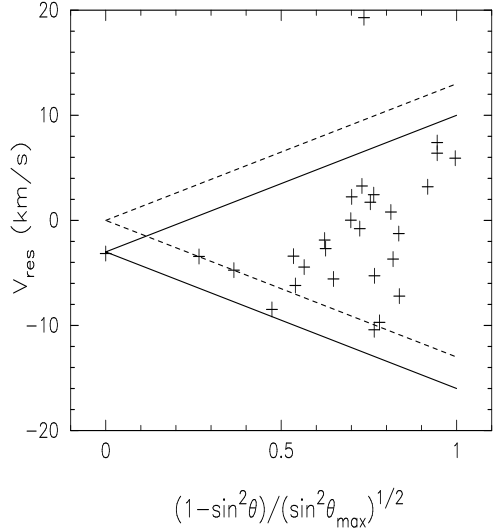,width=8cm,height=8cm,angle=-90}}
\end{center}
\caption[The residual radial velocity after removing the
Galactic rotation component plotted against the ``expansion
parameter'' {\it i.e.,} \mbox{$(1-sin^2\theta/sin^2\theta_{max})^{1/2}$}....]
{ The residual radial velocity after removing the contribution from
Galactic differential rotation plotted against the ``expansion
parameter'' {\it i.e.,} \mbox{$(1-sin^2\theta/sin^2\theta_{max})^{1/2}$}. $\theta$
is the angular separation of the source from the assumed center of expansion. If the
objects form an expanding shell, the points will lie on the two straight lines (the ``envelope'') shown. If
they are distributed in a volume, the points would tend to lie within the envelope, as is the case here. The
envelope in solid lines is for an expansion velocity 
of 13 km s$^{-1}$ with an offset of $-$3 km s$^{-1}$.
The points are better ``enclosed'' by the solid lines than the envelope defined by the dashed lines (no offset). }
\end{figure}
Our data is shown in figure~6. In this plot we have included only those
detections that are in the lower part of the shell (i.e. $b < 0$) so as to
avoid the confusing region near the galactic plane, for example, the Vela
Molecular Ridge. Although its estimated distance of $\sim 1\;{\rm kpc}$
would put the Ridge well beyond the region under study here, there will be 
blending in velocity space towards these longitudes. As may be seen in the
figure, the IRAS point sources (with which the molecular gas we have
detected is associated) are expanding about a common centre. The filled
nature of the cone suggests that the gas is not confined to a thin shell
but rather distributed over a volume, with the outer regions expanding
faster. From the slope of the two lines enveloping the data points we
deduce that the outer regions are expanding with a velocity $\sim
13\;\;{\rm km\;s}^{-1}$. The data also suggests that there may be offset
of $\sim -3\;\;{\rm km\;s}^{-1}$, i.e. the expansion is more {\it symmetric}
with respect to a residual radial velocity of $-3\;\;{\rm km\;s}^{-1}$
than about zero velocity. If this is significant then it would imply that
the gas has an overall drift. We shall return to this presently.

\subsection{The southern dark clouds} It may be recalled that our search
for molecular gas was motivated by a shell-like structure seen in the
distribution of IRAS point sources. Indeed, our candidates were a sample
of these point sources; due to the limited observing season at the site of
the telescope we used we could only observe about 100 of these sources.
Nevertheless, due to fortuitous circumstances, molecular observations of a
different class of objects -- some of which are possibly related to the
feature we were investigating -- became available to us. Recently a
general survey of the population of {\it dark clouds} in the southern sky
was undertaken in the $J = 1 \rightarrow 0$ line of $^{12}{\rm CO}$ using
the Mopra antenna (Otrupcek, Hartley and Wang Jing-Sheng, 1995). These
clouds listed in the Catalogue of Dark Clouds by Hartley {\it et al.}
(1986) appear as dark patches of obscuration in optical photographs.
Interestingly, the extinction map made from the ESO/SERC Southern Sky
Survey by Feitzinger and St\"{u}we (1984) shows a shell-like structure
just in the region of the IRAS shell. In view of this we decided to
enlarge our database of molecular gas in the region of the IRAS shell by
including the southern dark clouds. {\it The radial velocities from the Mopra
Survey were very kindly made available to us by Otrupcek.}

\begin{figure}[h]
\begin{center}
\mbox{
\epsfig{file=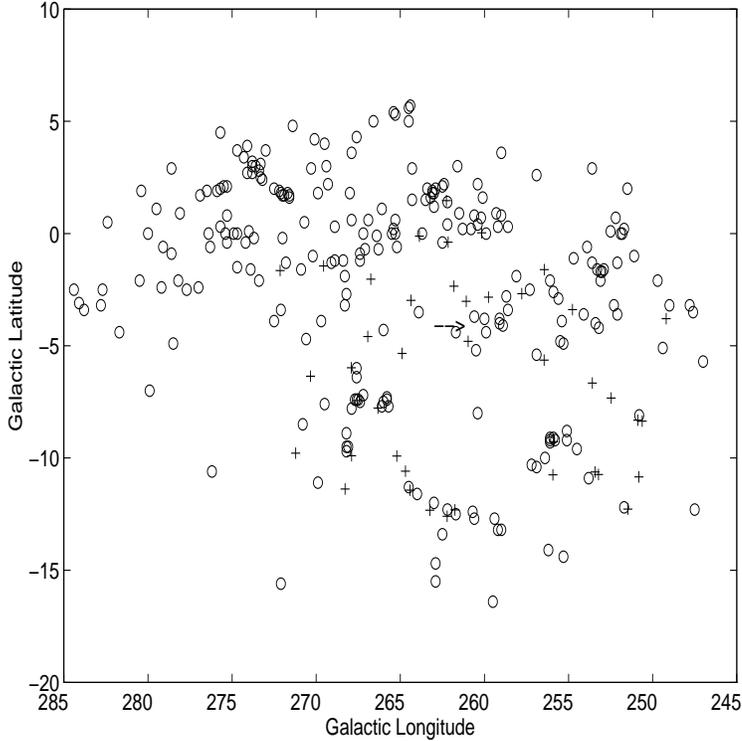,width=10cm,height=10cm}}
\end{center}
\caption[The molecular detections towards IRAS point
sources (plus signs), and the Southern Dark Clouds... ]
{The figure shows the molecular detections towards IRAS point
sources (plus signs), and the Southern Dark Clouds including cometary globules
in the region of the
IRAS Vela Shell shown as open circles. The arrow indicates the morphological center of
the cometary globule system.}
\end{figure}

In figure~7 we have replotted the molecular gas detected in this region,
this time including the dark clouds, as well as the cometary globules. Since
the molecular detections of the Soutern Dark Clouds is not yet published, we
have given in the Appendix an extensive Table of the coordinates and measured
LSR velocities of the subset of the Dark Clouds within the IRAS Vela Shell (i.e., within 12.5$^\circ$ from the assumed centre of expansion). The corresponding 
data for the cometary globules were taken from Sridharan (1992b).In figure~7, 
the
IRAS point sources observed by us are shown as plus signs, and the rest as
open circles. The tip of the {\it arrow} represents the point with respect
to which both the IRAS point sources as well as the cometary globules
appear to be expanding. Since the distribution of the dark clouds also
suggests a shell-like structure it is conceivable that they, too, are in a
state of expansion. This is indeed the case.
Figure~8 clearly shows that the molecular
clouds associated with the young stellar objects in the region, the
cometary globules, and a subset of the southern dark clouds are members of
a common family, and that they are expanding with respect to a common
centre.
\begin{figure}[h]
\begin{center}
\mbox{
\epsfig{file=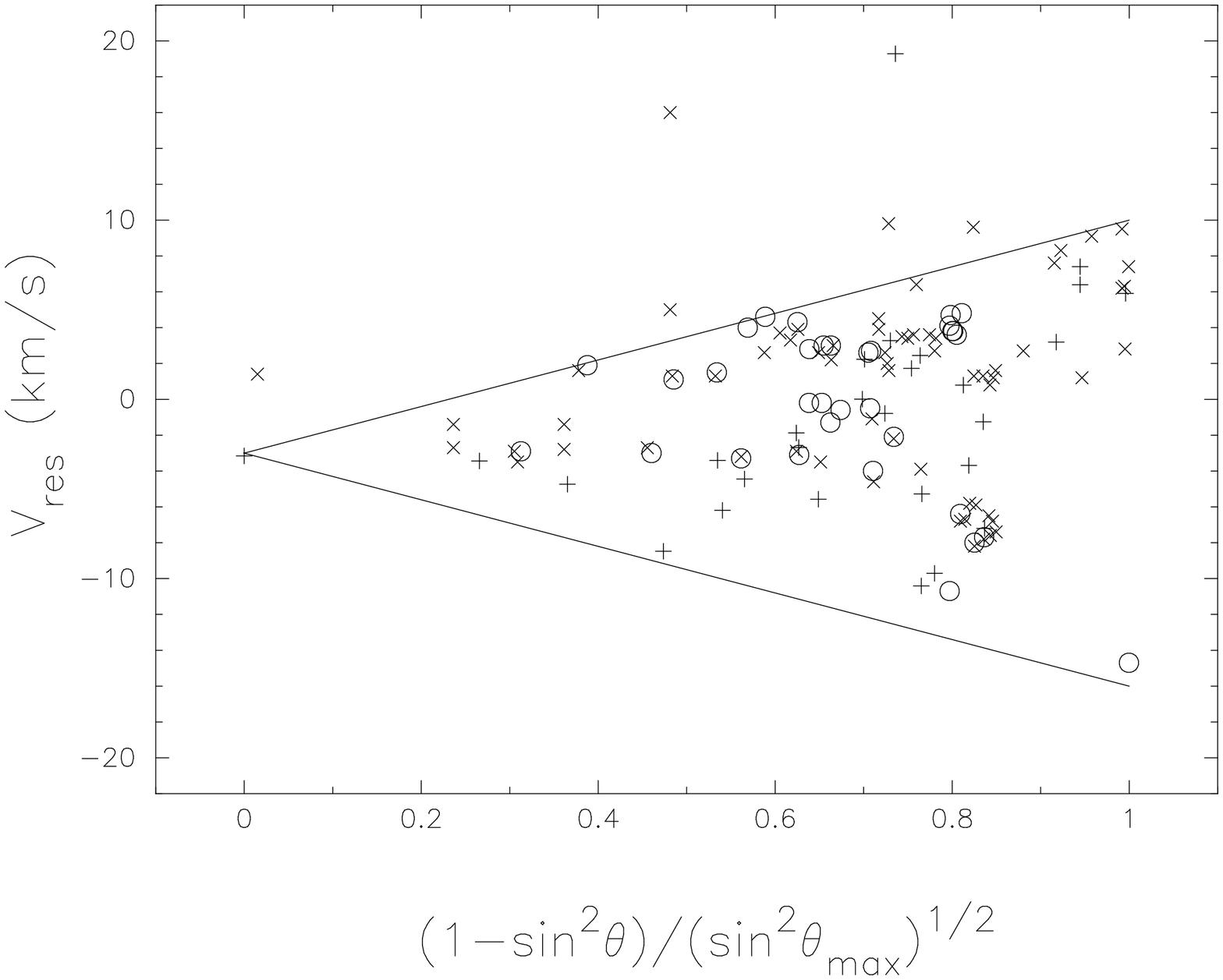,width=8cm,height=8cm,angle=-90}}
\end{center}
\caption{The residual velocity plotted against $(1-sin^2\theta/sin^2\theta_{max})^{1/2}$ for
{\em all} objects.
The crosses
denote SDCs, the open circles are cometary globules and the plus signs show molecular
detections towards IPSC sources. The ``expansion envelope'' is for an expansion velocity
of 13 \kms and offset of $-$3 \kms.}
\end{figure}
As in the analysis presented earlier in figure~6, to avoid confusion with unrelated
objects near the galactic plane we have included only those dark clouds
with latitudes greater than $-4^{\circ}$. We have, however, included all
the cometary globules; in view of their distinctive morphology there is
less chance of confusion. 
\subsection{A statistical test} To rule out the possibility that the
signature of expansion seen in figure~8 is spurious we did the following
test. It essentially involved {\it scrambling} the observed (residual)
radial velocities among the objects in the sample, and determining the
{\it fraction of points which lie within the envelope} defined by an
expansion velocity $v_{\rm exp}$ and an offset $v_{\rm offset}$. For every
assumed value of the `offset' we determined the minimum value of $v_{\rm
exp}$ for which $\geq 95\%$ of the observed points fell within the
expansion envelope. Given a pair $\left(v_{\rm exp}, v_{\rm
offset}\right)$ so determined, we generated a large set of random samples
and determined for each set the fraction of random samples that fall
within the wedge-shaped region in figure~9 . If the mean of this
distribution of fractions is not significantly different from the fraction
of the {\it actual} observed sample that lies within the defined envelope,
then the expansion deduced by us would not be statistically significant.
The most statistically significant values we obtained were $v_{\rm exp} =
13\;\;{\rm km\;s}^{-1}$ and $v_{\rm offset} = -3\;\;{\rm km\;s}^{-1}$. The
results of 400,000 simulations for this pair of values is shown in figure~9
. As may be seen, one can say with a confidence at $3.3 \sigma$ level that
the observed radial velocities (after correcting for contribution from
galactic rotation) indicates an expansion. For completeness we mention
that we also did simulations for offsets of 1, 0, $-2$ and $-4\;\;{\rm
km\;s}^{-1}$, all with $v_{\rm exp} = 13\;\;{\rm km\;s}^{-1}$. As
mentioned above, the most significant value for the offset in the residual
velocity was $-3\;\;{\rm km\;s}^{-1}$, with the significance decreasing on
either side of this value.
\begin{figure}[h]
\begin{center}
\mbox{
\epsfig{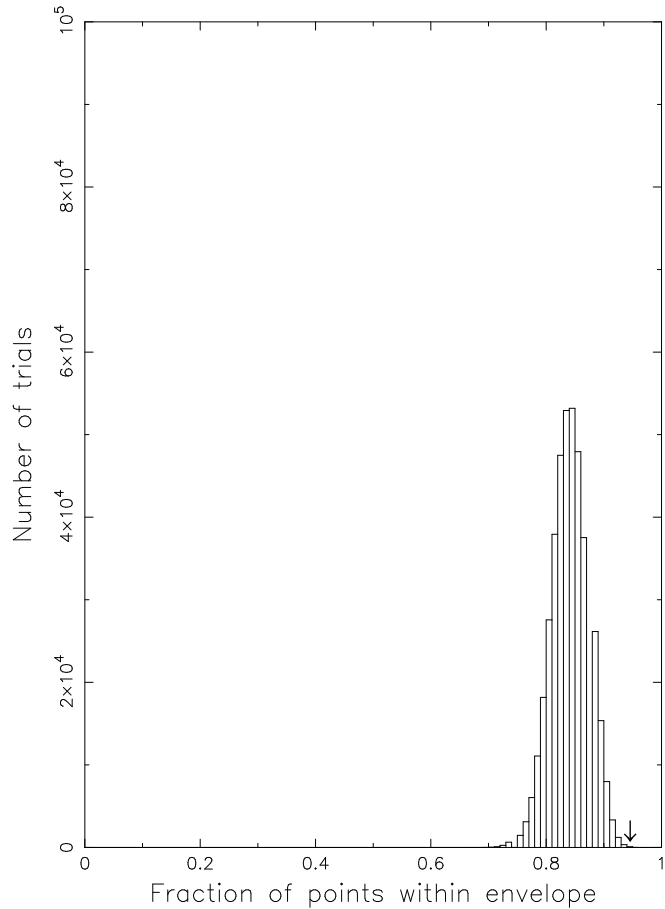}}
\end{center}
\caption[Result of significance tests done for an expansion velocity
of 13 km s$^{-1}$ and offset of $-$3 km s$^{-1}$.]{Result of significance tests done for an expansion velocity
of 13 km s$^{-1}$ and offset of $-$3 km s$^{-1}$. In these tests, the velocity axis in
the $V_{LSR}$ vs $(1-\sin^2\theta/\sin^2\theta_{max})^{1/2}$ diagram was ``scrambled'' and the number
of points within the ``expansion envelope'' computed for each trial. The peak of
the histogram shows the most likely fraction of points within the ``expansion envelope''
for a {\em random sample.} The arrow points to the fraction
of points within the ``expansion envelope'' for the {\em observed sample.}
This point lies 3.3 $\sigma$
away from the peak for the random samples.}
\end{figure}

\section{Summary and discussion} 
We first summarize our main conclusions:\\ \\
\noindent (i) There is a significant amount of molecular gas associated
with the IRAS point sources defining the shell-like feature. Our
observations thus confirm the expectation that molecular gas must be
associated with these Young Stellar Objects.\\ \\ (ii) Perhaps more
importantly, these point sources which delineate the ``IRAS Vela Shell''
are expanding about a common centre, with the sources in the outer region
moving with a velocity $\sim 13\;\;{\rm km\;s}^{-1}$.\\ \\ (iii) Our study
has established that a subset of the ``Southern Dark Clouds'' in this
region are also part of this ``shell'' since they, too, are participating
in the systematic motion mentioned above.\\ \\ (iv) Earlier observers had
found that the dozen or so cometary globules in this region, as well as
some ionized gas, showed evidence of similar expansion with roughly the
same velocity. One can therefore safely conclude that the cometary
globules and the expanding ionized gas are also part of the ``IRAS Vela
Shell''.

\subsection{The mass of the shell} 
We shall now attempt to estimate the
mass of the shell. From its infrared emissivity Sahu (1992) estimated the
amount of dust in the shell, and assuming the standard dust-to-gas ratio
she estimated the mass of the shell to be $\sim 10^{6} M_{\odot}$. As we
shall presently see, our estimates yield a mass an order of magnitude less
than this.

As already discussed, the lower half $(b < -4^{\circ})$ of the IRAS Vela
Shell is more clearly defined, and has an angular radius of $\sim
12.5^{\circ}$. Using the criteria explained earlier, we estimate that
there are $\sim 1000$ Young Stellar Objects in the lower half of the
shell. We shall now assume that these are of roughly the same mass as the
typical cometary globules. The mass of a typical globule has been
estimated to be $\leq 100 M_{\odot}$ (Sridharan 1992b, and references
therein). Adopting this value we estimate that the mass of the molecular
gas associated with the entire IRAS Vela Shell must be $\sim 10^{5}
M_{\odot}$. In deriving this estimate we have allowed for the fact that we
detected molecular gas in only $\sim 50\%$ of the IPSC sources towards
which we looked.

An alternative mass estimate can be made as follows. It has been argued
that in the local giant molecular cloud complexes such as Orion, Ophiuchus
and Taurus-Auriga the {\it efficiency} with which gas is converted to
stars is $\sim 1\%$ (Evans and Lada, 1990). If this is also the case in the
small molecular clouds under discussion, and if the typical mass of
the stars formed in these clouds is $\sim 1 M_{\odot}$, then the presence
of approximately 1000 Young Stellar Objects in the shell suggests a total
mass $\sim 10^{5} M_{\odot}$, consistent with the earlier estimate. But if
the star forming efficiency is much higher in the small globules, as has
been argued, for example, by Bhatt (1993), then the mass of the molecular gas could be
smaller.

To derive the total mass of the shell one must, of course, add the mass of
the ionized gas, as well as neutral atomic gas associated with the shell.
While there is clear evidence for some ionized gas (HII, NII, SII etc.) it
is difficult to estimate its mass. As for HI associated with the Vela
Shell, the picture is far from clear.

\subsection{On the origin and evolution of the IRAS Vela Shell}
Based on
the fact that the Vela OB2 association of stars appears centrally located
with respect to the shell, Sahu (1992) attributed the expansion of the gas
in the shell to the combined effect of stellar wind from the massive stars
and supernova explosions in the association. Sridharan (1992a) invoked the same explanation to
account for the expansion of the system of cometary globules. We endorse
these suggestions, and our observations lend more credence to the scenario
that the
Vela Shell is the remnant of the giant molecular cloud from which the Vela
OB2 association itself formed.

Although the Vela OB2 association is more or less symmetrically located
with respect to the IRAS Vela Shell, there are two points to consider: (i)
whether the group of stars are members of a genuine ``association'', and
(ii) whether the shell and the association are at the same distance from
us. As for the first point, the recent proper motion measurements by the
Hipparcos satellite firmly establishes this group of stars as a genuine
association with 116 members, including the O star $\gamma^{2}$ Velorum, at
a mean distance of $415 \pm 10\;\;{\rm pc}$ (de~Zeeuw {\it et al.}, 1997).
Hipparcos data also lends support to the idea that it is a fairly evolved
association approximately $10^{7}$ years old (Schaerer, Schmutz and
Grenon, 1997). In comparison, the distance estimate to the shell is
indirect. Earlier we presented arguments to support the hypothesis that
the system of cometary globules in the region is part of the expanding
shell. The estimated distances to a number of these globules roughly
agrees with the Hipparcos distance to the Vela OB2 association (Brand {\it et al.},
1983; Pettersson, 1987). There is also a piece of circumstantial evidence
which may be more reliable. A recent photometric study of the $H\alpha$
emission from the bright rim of the globule CG22 (Rajagopal, 1997) clearly
indicates that the ionizing source is $\zeta$-Puppis.
The Hipparcos measurements yield a
distance of $429^{+120}_{-77}\;{\rm pc}$ to $\zeta$-Puppis 
(van der Hucht {\it et al.}, 1997). This would
suggest that the distance to CG22 is of this order, and consistent with
the distance to the Vela OB2 association.

If one accepts the premise that the Vela OB2 association and the IRAS Vela
Shell are at the same distance, then one has to ask if it is plausible
that the expansion of the shell is causally connected with the group of
stars. Adapting the model due to McCray and Kafatos (1987) for the
formation of ``supershells'' by OB associations, Sahu (1992) has argued
that if Vela OB2 is a ``standard association'' of the type found within
1~kpc of the Sun then it could account for the observed expansion with a
kinetic energy $\sim 10^{50} - 10^{51}\;{\rm erg}$. A closer examination
of this important question is warranted in the light of the Hipparcos
observation, and we hope to undertake it.

There is an increasing body of evidence to suggest that the break up of
giant molecular clouds may be quite common. Many of the
giant molecular clouds in our neighbourhood show evidence of streaming
flows of ionized gas, clumpy distribution of the molecular gas, large
velocity fields etc.. These phenomena are consistent with these giant
clouds disintegrating under the influence of nearby OB associations
(Leisawitz, Bash and Thaddeus, 1989). Such clouds may represent an {\it
earlier} stage of evolution of the IRAS Vela Shell. As for its future
evolution, it is conceivable that as the expanding molecular gas sweeps up
enough interstellar matter it will develop into a classical
``supershell''.
\section*{Acknowledgements :}
This study was undertaken at the suggestion of 
Prof. A. Blaauw. We wish to thank him for his continued interest,
as well as critical comments.
We also thank T.K. Sridharan for his help
at various stages. The data from the Mopra survey was invaluable;
we are indebted to R. Otrupcek and co-workers for giving us timely
access to it. Finally, the support from the staff at the Raman
Institute millimeter wave observatory is gratefully acknowledged.
\newpage
{}

\section*{Appendix}
In the following set of tables, we present the Galactic co-ordinates, measured velocities (LSR) of the
$^{12}$CO emission, and projected separation from the assumed center of the  Shell for the Southern Dark Clouds (SDCs). Only those SDCs within 12.5$^\circ$ of
the center have been included.

\begin{table}
\begin{center}
\begin{tabular}{|l|l|l|l|||l|l|l|l|}\hline
l$^{\rm II}$ & b$^{\rm II}$ & V$_{lsr}$ & $\theta$& l$^{\rm II}$ & b$^{\rm II}$ & V$_{lsr}$ & $\theta$\\ \hline\hline
 Deg  & Deg &  km/s & Deg& Deg  & Deg &  km/s & Deg\\ \hline
247.80 &   -3.2 &   18.7 &  12.40 & 	253.40 &   -4.0 &   11.8 &   6.77 \\ 
249.00 &   -3.2 &   17.2 &  11.20 & 	253.60 &   -1.3 &   11.2 &   7.12 \\ 
249.00 &   -3.2 &   26.2 &  11.20 & 	253.60 &    2.9 &    5.9 &   9.57 \\ 
249.40 &   -5.1 &    7.2 &  10.82 & 	253.80 &  -10.9 &   -1.3 &   9.36 \\ 
249.40 &   -5.1 &   18.2 &  10.82 & 	253.90 &   -0.6 &   10.8 &   7.16 \\ 
249.70 &   -2.1 &   15.6 &  10.65 & 	253.90 &   -0.6 &   35.8 &   7.16 \\ 
250.80 &   -8.1 &   -1.0 &  10.21 & 	254.10 &   -3.6 &   10.6 &   6.09 \\ 
251.10 &   -1.0 &    9.6 &   9.57 & 	254.50 &   -9.6 &   -1.7 &   7.94 \\ 
251.50 &    2.0 &   25.1 &  10.57 & 	254.70 &   -1.1 &    9.8 &   6.22 \\ 
251.70 &    0.2 &    5.1 &   9.48 & 	254.70 &   -1.1 &   37.1 &   6.22 \\ 
251.70 &  -12.2 &   -1.3 &  11.76 & 	255.10 &   -9.2 &   -4.6 &   7.23 \\ 
251.80 &    0.0 &    5.6 &   9.30 & 	255.10 &   -8.8 &   -6.0 &   6.95 \\ 
251.90 &    0.0 &    5.4 &   9.21 & 	255.30 &  -14.4 &    3.8 &  11.44 \\ 
252.10 &   -3.6 &   -2.0 &   8.08 & 	255.30 &   -4.9 &    9.8 &   4.95 \\ 
252.10 &   -3.6 &    7.2 &   8.08 & 	255.40 &   -3.9 &    9.8 &   4.77 \\ 
252.10 &   -3.6 &   14.5 &   8.08 & 	255.50 &   -4.8 &   10.5 &   4.73 \\ 
252.10 &   -1.3 &   12.9 &   8.53 & 	255.60 &   -2.9 &    8.9 &   4.71 \\ 
252.20 &    0.7 &    1.6 &   9.28 & 	255.80 &   -9.2 &   -5.6 &   6.76 \\ 
252.30 &   -3.2 &    8.5 &   7.92 & 	255.90 &   -9.1 &   -5.4 &   6.62 \\ 
252.30 &   -3.2 &   12.5 &   7.92 & 	255.90 &   -2.6 &    9.8 &   4.51 \\ 
252.50 &    0.1 &    4.4 &   8.72 & 	256.10 &   -2.1 &    9.0 &   4.51 \\ 
252.90 &   -1.6 &    5.9 &   7.67 & 	256.10 &   -9.3 &   -4.3 &   6.64 \\ 
252.90 &   -1.6 &   11.2 &   7.67 & 	256.10 &   -9.2 &   -4.6 &   6.57 \\ 
253.00 &   -1.7 &    5.7 &   7.55 & 	256.10 &   -9.1 &   -5.2 &   6.49 \\ 
253.00 &   -1.7 &   10.3 &   7.55 & 	256.20 &  -14.1 &    3.5 &  10.81 \\ 
253.10 &   -2.1 &   11.6 &   7.34 & 	256.40 &  -10.0 &   -3.6 &   7.05 \\ 
253.10 &   -1.7 &    5.4 &   7.45 & 	256.90 &  -10.4 &   -4.5 &   7.14 \\ 
253.10 &   -1.7 &    9.9 &   7.45 & 	256.90 &   -5.4 &   11.3 &   3.54 \\ 
253.20 &   -4.2 &   11.8 &   6.97 & 	256.90 &    2.6 &   10.6 &   7.41 \\ 
253.30 &   -1.6 &    5.2 &   7.30 & 	257.20 &  -10.3 &   -3.7 &   6.92 \\ \hline
\end{tabular}
\caption[Southern Dark Clouds within 12.5$^\circ$ of the assumed center of the IRAS Vela Shell]
{Southern Dark Clouds within 12.5$^\circ$ of the center of the IRAS Vela Shell, from the Mopra
Survey (Otrupcek \etal, 1995) kindly made available to us by R.Otrupcek.: Column 1 and 2 
give the galactic co-ordinates, Column 3
shows the LSR velocity and
Column 4 shows the projected angular separation between the source and the center of the Shell. {\bf The table
is continued in Columns 5 to 8.}}
\end{center}
\end{table}

\begin{table}
\begin{center}
\begin{tabular}{|l|l|l|l|||l|l|l|l|}\hline
l$^{\rm II}$ & b$^{\rm II}$ & V$_{lsr}$ & $\theta$& l$^{\rm II}$ & b$^{\rm II}$ & V$_{lsr}$ & $\theta$\\ \hline\hline
 Deg  & Deg &  km/s & Deg& Deg  & Deg &  km/s & Deg\\ \hline
257.30 &   -2.5 &   15.9 &   3.26 & 	260.40 &    0.4 &    6.3 &   4.46 \\ 
258.10 &   -1.9 &   10.3 &   2.99 & 	260.40 &    0.4 &    9.2 &   4.46 \\ 
258.10 &   -1.9 &   43.2 &   2.99 & 	260.40 &   -8.0 &    3.4 &   3.96 \\ 
258.60 &    0.3 &    7.9 &   4.63 & 	260.40 &    2.2 &    7.5 &   6.25 \\ 
258.60 &    0.3 &   17.3 &   4.63 & 	260.50 &   -5.2 &    5.0 &   1.20 \\ 
258.60 &   -3.4 &    8.9 &   1.70 & 	260.60 &    0.8 &    5.2 &   4.87 \\ 
258.60 &   -3.4 &   15.1 &   1.70 & 	260.60 &   -3.7 &    4.5 &   0.55 \\ 
258.70 &   -2.8 &   10.3 &   1.93 & 	260.60 &   -3.7 &    6.4 &   0.55 \\ 
258.90 &   -4.1 &    8.5 &   1.27 & 	260.60 &  -12.7 &   -2.4 &   8.66 \\ 
259.00 &  -13.2 &    4.4 &   9.23 & 	260.70 &  -12.4 &    0.0 &   8.37 \\ 
259.00 &    0.8 &    5.9 &   4.99 & 	260.80 &    0.2 &    5.8 &   4.30 \\ 
259.00 &    3.6 &    2.6 &   7.74 & 	260.80 &    0.2 &    8.0 &   4.30 \\ 
259.00 &    3.6 &    6.4 &   7.74 & 	260.80 &    0.2 &   12.2 &   4.30 \\ 
259.10 &   -4.0 &    9.5 &   1.07 & 	261.30 &    0.2 &    5.7 &   4.40 \\ 
259.10 &   -3.8 &    9.5 &   1.10 & 	261.30 &    0.2 &    7.6 &   4.40 \\ 
259.20 &    0.3 &    4.4 &   4.46 & 	261.30 &    0.2 &   10.9 &   4.40 \\ 
259.20 &    0.3 &    6.5 &   4.46 & 	261.50 &    0.9 &    6.4 &   5.12 \\ 
259.20 &    0.3 &   12.3 &   4.46 & 	261.60 &    3.0 &   11.2 &   7.19 \\ 
259.20 &  -13.2 &    5.2 &   9.20 & 	261.70 &   -4.4 &    8.4 &   1.57 \\ 
259.30 &    0.9 &    3.6 &   5.03 & 	261.70 &   -4.4 &   11.7 &   1.57 \\ 
259.30 &    0.9 &    6.7 &   5.03 & 	261.70 &  -12.5 &    6.1 &   8.59 \\ 
259.30 &    0.9 &    4.5 &   5.03 & 	262.20 &    0.4 &    8.4 &   4.89 \\ 
259.40 &  -12.7 &    1.1 &   8.68 & 	262.20 &  -12.3 &    4.8 &   8.50 \\ 
259.50 &  -16.4 &    3.6 &  12.37 & 	262.20 &    1.4 &    4.8 &   5.81 \\ 
259.90 &   -4.4 &    9.6 &   0.44 & 	262.40 &    2.2 &    6.5 &   6.63 \\ 
259.90 &    0.0 &    7.9 &   4.06 & 	262.40 &    2.2 &    8.4 &   6.63 \\ 
260.00 &   -3.8 &  -12.7 &   0.30 & 	262.50 &  -13.4 &   -0.7 &   9.64 \\ 
260.10 &    1.6 &    5.6 &   5.65 & 	262.50 &   -0.4 &    7.6 &   4.33 \\ 
260.10 &    1.6 &    6.8 &   5.65 & 	262.50 &   -0.4 &   39.3 &   4.33 \\ 
260.20 &    0.7 &    5.3 &   4.75 & 	262.50 &    2.1 &    8.4 &   6.58 \\ \hline
\end{tabular}
\setcounter{table}{1}
\caption[continued from the previous page: Southern Dark Clouds within 12.5$^\circ$ of...]
{continued from the previous page: Southern Dark Clouds within 12.5$^\circ$ of the assumed center of the
IRAS Vela Shell, from the Mopra
Survey (Otrupcek \etal, 1995): Column 1 and 2 give the galactic co-ordinates, Column 3
shows the LSR velocity and
Column 4 shows the projected angular separation between the source and the center of the Shell. {\bf The table
is continued in Columns 5 to 8.}}
\end{center}
\end{table}

\begin{table}
\begin{center}
\begin{tabular}{|l|l|l|l|||l|l|l|l|}\hline
l$^{\rm II}$ & b$^{\rm II}$ & V$_{lsr}$ & $\theta$& l$^{\rm II}$ & b$^{\rm II}$ & V$_{lsr}$ & $\theta$\\ \hline\hline
 Deg  & Deg &  km/s & Deg& Deg  & Deg &  km/s & Deg\\ \hline
262.90 &  -14.7 &   -0.5 &  10.99 & 	265.50 &    0.0 &   -6.8 &   6.69 \\ 
262.90 &    2.0 &    5.0 &   6.64 & 	265.70 &   -7.7 &    3.0 &   6.62 \\ 
263.00 &    1.8 &    7.1 &   6.50 & 	265.80 &   -7.3 &    3.8 &   6.50 \\ 
263.00 &    1.8 &   26.9 &   6.50 & 	265.80 &   -7.4 &    3.4 &   6.55 \\ 
263.00 &    1.2 &    7.8 &   5.96 & 	266.00 &   -7.5 &    3.5 &   6.77 \\ 
263.00 &  -12.0 &    3.8 &   8.44 & 	266.00 &   -4.3 &    4.9 &   5.83 \\ 
263.10 &    1.8 &    5.6 &   6.54 & 	266.10 &    1.1 &    2.1 &   7.85 \\ 
263.10 &    1.9 &    3.8 &   6.63 & 	266.10 &    1.1 &    5.4 &   7.85 \\ 
263.20 &    1.6 &    5.4 &   6.41 & 	266.10 &    1.1 &    8.4 &   7.85 \\ 
263.40 &    2.0 &    3.0 &   6.86 & 	266.10 &    1.1 &   12.0 &   7.85 \\ 
263.50 &    1.5 &    4.8 &   6.47 & 	266.10 &    1.1 &   21.6 &   7.85 \\ 
263.50 &    1.5 &    8.7 &   6.47 & 	266.10 &   -7.7 &    3.5 &   6.96 \\ 
263.70 &    0.0 &    5.4 &   5.37 & 	266.30 &   -0.7 &    3.8 &   6.98 \\ 
263.70 &    0.0 &    9.9 &   5.37 & 	266.30 &   -0.7 &    6.2 &   6.98 \\ 
263.90 &   -3.5 &    2.8 &   3.77 & 	266.30 &   -0.7 &   10.0 &   6.98 \\ 
263.90 &   -3.5 &    5.2 &   3.77 & 	266.40 &   -0.1 &    3.8 &   7.37 \\ 
264.00 &  -11.6 &    4.2 &   8.47 & 	266.40 &   -0.1 &    5.7 &   7.37 \\ 
264.30 &    2.9 &   11.3 &   8.08 & 	266.60 &    5.0 &    2.5 &  11.10 \\ 
264.30 &    1.5 &    6.6 &   6.92 & 	266.90 &    0.6 &    5.3 &   8.18 \\ 
264.40 &    5.7 &    8.9 &  10.63 & 	267.10 &   -0.7 &    1.4 &   7.70 \\ 
264.50 &    5.6 &   10.0 &  10.58 & 	267.10 &   -0.7 &    6.0 &   7.70 \\ 
264.50 &    5.0 &    2.5 &  10.03 & 	267.10 &   -0.7 &   18.8 &   7.70 \\ 
264.50 &  -11.3 &   12.0 &   8.44 & 	267.20 &   -7.2 &    4.9 &   7.70 \\ 
265.20 &   -0.6 &    5.7 &   6.10 & 	267.20 &    0.0 &    4.5 &   8.11 \\ 
265.30 &    0.0 &    6.5 &   6.53 & 	267.20 &    0.0 &    6.7 &   8.11 \\ 
265.30 &    5.3 &    3.0 &  10.66 & 	267.20 &    0.0 &    8.8 &   8.11 \\ 
265.30 &    0.6 &    4.7 &   6.92 & 	267.40 &   -0.9 &    4.0 &   7.88 \\ 
265.30 &    0.6 &    6.7 &   6.92 & 	267.40 &   -0.9 &    6.6 &   7.88 \\ 
265.40 &    0.2 &    6.0 &   6.74 & 	267.40 &   -0.9 &    8.3 &   7.88 \\ 
265.40 &    0.2 &    7.7 &   6.74 & 	267.40 &   -7.5 &    8.6 &   8.01 \\ \hline
\end{tabular}
\setcounter{table}{1}
\caption[continued from the previous page: Southern Dark Clouds within 12.5$^\circ$ of..]
{continued from the previous page: Southern Dark Clouds within 12.5$^\circ$ of the assumed center of the
IRAS Vela Shell, from the Mopra
Survey (Otrupcek \etal, 1995): Column 1 and 2 give the galactic co-ordinates, Column 3
shows the LSR velocity and
Column 4 shows the projected angular separation between the source and the center of the Shell. {\bf The table
is continued in Columns 5 to 8.}}
\end{center}
\end{table}

\begin{table}
\begin{center}
\begin{tabular}{|l|l|l|l|||l|l|l|l|}\hline
l$^{\rm II}$ & b$^{\rm II}$ & V$_{lsr}$ & $\theta$& l$^{\rm II}$ & b$^{\rm II}$ & V$_{lsr}$ & $\theta$\\ \hline\hline
 Deg  & Deg &  km/s & Deg& Deg  & Deg &  km/s & Deg\\ \hline
267.50 &   -7.4 &    5.8 &   8.06 & 	269.40 &    3.0 &   -1.4 &  11.61 \\ 
267.60 &   -6.4 &    5.8 &   7.79 & 	269.50 &   -7.6 &    4.8 &   9.98 \\ 
267.60 &   -6.0 &    5.6 &   7.68 & 	269.50 &    4.0 &   -3.8 &  12.32 \\ 
267.60 &    4.3 &    1.3 &  11.18 & 	269.70 &   -3.9 &    2.2 &   9.53 \\ 
267.60 &   -7.4 &    5.6 &   8.15 & 	269.90 &  -11.1 &   -0.5 &  12.01 \\ 
267.70 &   -7.4 &    5.7 &   8.24 & 	269.90 &  -11.1 &    0.8 &  12.01 \\ 
267.90 &    0.6 &    0.1 &   9.02 & 	269.90 &    1.8 &   -3.0 &  11.35 \\ 
267.90 &    0.6 &    2.7 &   9.02 & 	269.90 &    1.8 &    4.2 &  11.35 \\ 
267.90 &    3.6 &    0.5 &  10.87 & 	270.20 &   -1.0 &    1.8 &  10.48 \\ 
267.90 &    3.6 &    2.1 &  10.87 & 	270.20 &   -1.0 &    7.2 &  10.48 \\ 
267.90 &   -7.8 &    6.7 &   8.59 & 	270.20 &   -1.0 &    9.6 &  10.48 \\ 
268.00 &    1.8 &   -5.4 &   9.77 & 	270.30 &    2.9 &   -4.2 &  12.28 \\ 
268.00 &    1.8 &   -0.6 &   9.77 & 	270.60 &   -4.7 &    3.5 &  10.45 \\ 
268.10 &   -9.5 &    6.1 &   9.62 & 	270.70 &    0.5 &   -0.2 &  11.47 \\ 
268.20 &   -9.7 &    5.9 &   9.82 & 	270.70 &    0.5 &    6.2 &  11.47 \\ 
268.20 &   -8.9 &    4.8 &   9.38 & 	270.80 &   -8.5 &   -0.6 &  11.52 \\ 
268.20 &   -9.5 &    5.5 &   9.70 & 	270.80 &   -8.5 &    0.8 &  11.52 \\ 
268.20 &   -2.7 &    7.5 &   8.14 & 	270.90 &   -1.6 &    5.6 &  11.00 \\ 
268.20 &   -2.7 &   14.4 &   8.14 & 	271.80 &   -1.3 &   -3.9 &  11.95 \\ 
268.30 &   -3.2 &    1.1 &   8.17 & 	271.80 &   -1.3 &   -1.4 &  11.95 \\ 
268.30 &   -1.9 &    3.1 &   8.41 & 	271.80 &   -1.3 &    2.7 &  11.95 \\ 
268.40 &   -1.2 &    4.7 &   8.71 & 	272.00 &   -0.2 &   -0.8 &  12.44 \\ 
268.90 &    0.3 &    0.7 &   9.75 & 	272.00 &   -0.2 &    4.8 &  12.44 \\ 
268.90 &    0.3 &    4.8 &   9.75 & 	272.10 &   -3.4 &    8.2 &  11.95 \\ 
268.90 &    0.3 &    8.4 &   9.75 & 	272.50 &   -3.9 &    4.8 &  12.33 \\ 
268.90 &   -1.2 &    4.7 &   9.18 & 	272.50 &   -3.9 &    6.2 &  12.33 \\ 
269.10 &   -1.3 &    2.6 &   9.34 & 	       &        &        &        \\
269.10 &   -1.3 &    6.0 &   9.34 & 	       &        &        &        \\
269.10 &   -1.3 &    7.7 &   9.34 & 	       &        &        &        \\
269.10 &   -1.3 &    9.1 &   9.34 & 	       &        &        &        \\ \hline
\end{tabular}
\setcounter{table}{1}
\caption[continued from the previous page: Southern Dark Clouds within 12.5$^\circ$ of ...]
{continued from the previous page: Southern Dark Clouds within 12.5$^\circ$ of the assumed center of the
IRAS Vela Shell, from the Mopra
Survey (Otrupcek \etal, 1995): Column 1 and 2 give the galactic co-ordinates, Column 3
shows the LSR velocity and
Column 4 shows the projected angular separation between the source and the center of the Shell. {\bf The
table is continued in columns 5 to 8.}}
\end{center}
\end{table}

\end{document}